\documentclass[PRL,twocolumn,amsmath,amsfont,amssymb,floatfix]{revtex4}

\usepackage[]{graphicx}
\usepackage[]{tabularx}
\usepackage[]{psfrag}
\usepackage[]{textcomp}

\begin{document}

\title{Losses in coplanar waveguide resonators at millikelvin temperatures}

\author{P. Macha}
\email[pascal.macha@ipht-jena.de]{}
\author{S.H.W. van der Ploeg}
\author{G. Oelsner}
\author{E. Il'ichev}
\author{H.-G. Meyer}
\affiliation{Institute of Photonic Technology, PO Box 100239, D-07702 Jena, Germany}
\author{S. W\"unsch}
\author{M. Siegel}
\affiliation{Universit\"at Karlsruhe, Institut f\"ur Mikro-- und Nanoelektronische Systeme,
Hertzstra\ss e 16, D-76187 Karlsruhe, Germany}
\date{\today}

\begin{abstract}We study the loss rate for a set of $\lambda / 2$ coplanar waveguide resonators at millikelvin temperatures ($20\;\text{mK}$ -- $900\;\text{mK}$) and different applied powers ($3\cdot10^{-19}\;\text{W}$ -- $10^{-12}\;\text{W}$). The loss rate becomes power independent below a critical power. For a fixed power, the loss rate increases significantly with decreasing temperature. We show that this behavior can be caused by two-level systems in the surrounding dielectric materials. Interestingly, the influence of the two-level systems is of the same order of magnitude for the different material combinations. That leads to the assumption that the nature of these two-level systems is material independent.
\end{abstract}

\maketitle

 \psfrag{Nb-Al2O3}[Bl]{Nb-Al$_2$O$_3$}
 \psfrag{Al-Al2O3}[Bl]{Al-Al$_2$O$_3$}
 \psfrag{Nb-Si}[Bl]{Nb-Si   }

The interest in the properties of coplanar waveguide  resonators (CPWR) at low temperatures is caused by their potential application for quantum information processing devices \cite{Wallraff2004,DiCarlo2009} and photon detection down to the quantum limit \cite{Day2003}. In practice, for quantum--limited measurements or for preserving coherence in quantum circuits with CPWRs, the loss rate of the resonators should be minimized. The properties, especially the resonance frequency and noise behavior --- of CPWRs at low temperatures are known to be dominated by two--level systems (TLS) present in the active region of the resonator \cite{Gao2008a,Kumar2008,Barends2008}. These TLSs cause decoherence in mesoscopic quantum system \cite{Martinis2005} and lead to increased losses in CPWRs \cite{Lindstrom2009,Wang2009}. In this letter we analyze different resonator and substrate materials in order to find optimal conditions for quantum measurements on the chip and to give a description of TLSs induced losses in CPWRs. Therefore oxides and nitrides which are known sources of TLSs were as far as possible avoided and high resisitivity substrates were used.

\begin{figure}
 \psfrag{s}[Bl]{Superconductor}
 \psfrag{c}[Bl][bl][0.9]{Central conductor}
 \psfrag{g}[Bl][bl][0.9]{Ground plates}
 \psfrag{a}[Bl][bl][0.9]{Capacitance}
 \psfrag{u}[Bl]{Substrate}
\includegraphics[width=0.90\linewidth]{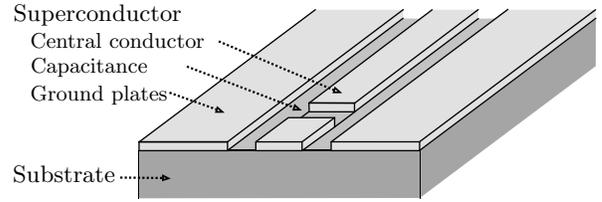}%
\caption{\label{fig:Scheme}The scheme of the the geometry of the CPWRs (not to scale). In order to avoid losses due to meanders the central conductor is a straight line. The width of the central conductor is $50\;$\textmu m, and the gap between the central conductor and the ground plate is $30\;$\textmu m which results in a impedance of $50\;\Omega$. The resonator is defined by two capacitances with a width of $90\;$\textmu m in the central conductor of CPWR to the input and output transmission line. The resonators are designed to be under--coupled.}
\end{figure}

Thus motivated by material science, we did not use any buffer layer or top dielectric layers for the resonators fabrication.  The superconducting material was directly deposited to the dielectric substrates and CPWRs of identical geometry were fabricated (see Fig. \ref{fig:Scheme}). We have investigated three resonators, namely aluminum (Al) and niobium (Nb) on sapphire as a substrate and niobium on undoped silicon (Si) as  a substrate ($>5 \text{k}\Omega \text{cm}$). The CPWRs were fabricated by dry etching of a $220$--nm--thick Nb film and by lifting off a $250$--nm--thick Al film respectively. The frequency of the fundamental mode of this half--wavelength transmission--type CPWR is $2.5\;\text{GHz}$ with Si as a substrate and $2.7\;\text{GHz}$ with sapphire respectively.

Two cryoperm shields and one superconducting shield made of lead enclosed the resonators in order to provide a stable environment and to avoid effects from external magnetic fields \cite{Healey2008, Song2008}. The samples were thermally anchored to the mixing chamber of a dilution refrigerator. The amplitude of the microwave transmission has been measured as a function of temperature and applied microwave power. The resonance frequency $\nu_0$ and spectral width $\Delta \nu_0$ of the resonator were extracted from the Lorentzian shape of the resonance curve. The signal transmitted through the resonator was amplified using a cryogenic amplifier at the $4.2\;\text{K}$ stage and room--temperature amplifiers.

The resonance frequency $\nu_0$ and spectral width $\Delta \nu_0$ which is connected to the overall loss rate $\delta = 2 \pi \Delta \nu_0$ (in the sense of a photon decay rate) correspond to the (loaded) quality factor $Q=2\pi\nu_0/\delta$. In our case, for under--coupled resonators, the overall loss rate is dominated by the intrinsic loss mechanisms of the resonator.  The loss rate is a function of temperature $T$ and applied power $P$ whereas the resonance frequency only depends on temperature. Above a temperature of about $T_c/10$ (about $1\;$K for Nb and $0.2\;$K for Al; $T_c$ is the critical temperature) the behavior of the resonator is dominated by the number of quasi--particles which decreases with temperature. Subsequently the loss rate decreases and converges to a minimum at $T_c/10$. The corresponding quality factors are $5.7\cdot10^{5}$ for Nb on Si, $4.5\cdot10^{5}$ for Nb on sapphire and $3.6\cdot10^{5}$ for Al on sapphire. This work centers on the behavior below $T_c/10$ where the number of quasi--particles is negligible and quantum effects become visible.

\begin{figure}
 \psfragscanon
 \psfrag{xl}[c]{Power at resonator input $P$ (W)}
 \psfrag{yl}[c]{Loss rate $\delta$ (kHz)}
 \psfrag{-18}{$10^{-18}$} \psfrag{-16}{$10^{-16}$} \psfrag{-14}{$10^{-14}$}\psfrag{-12}{$10^{-12}$}
 \psfrag{40}{40} \psfrag{60}{60} \psfrag{80}{80}  \psfrag{100}{100}   \psfrag{120}{120}
\includegraphics[width=0.90\linewidth]{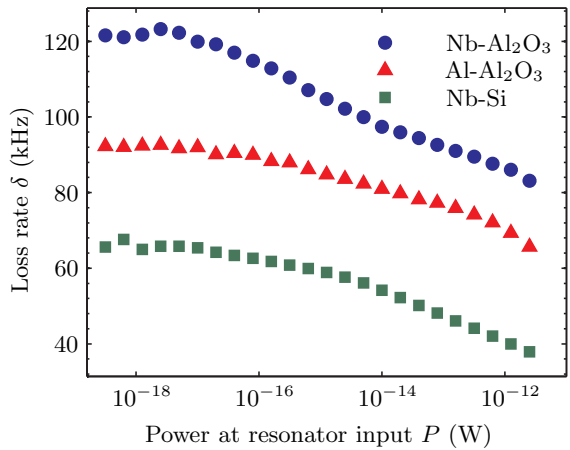}%
\caption{\label{fig:PowerOverview}The power dependence of the loss rate at a temperature of $20$ mK for the three different samples.}
\end{figure}

\begin{figure}
 \psfragscanon
 \psfrag{xl}[c]{Temperature $T$ (mK)}
 \psfrag{yl}[c]{Loss rate $\delta$ (kHz)}
 \psfrag{40}{40} \psfrag{60}{60} \psfrag{80}{80}  \psfrag{100}{100}   \psfrag{120}{120}
 \psfrag{200}{200} \psfrag{400}{400} \psfrag{600}{600}  \psfrag{800}{800}
\includegraphics[width=0.90\linewidth]{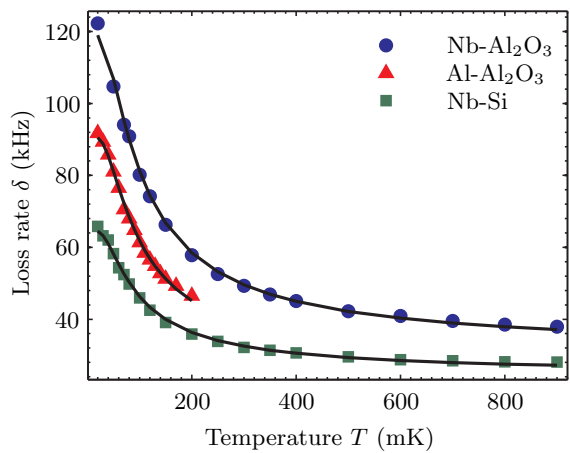}%
\caption{\label{fig:LowPowerLimit}The temperature dependence of the loss rate at the low power limit. Solid lines show fit according to Eq. (\ref{eq:Model}).}
\end{figure}

In order to quantify the function $\delta(T,P)$ in the region of interest let us qualitatively discuss its power dependence at fixed temperature. For high enough power the response of the CPWR is non--linear \cite{Abdo2006}. In the linear regime, the loss rate $\delta$ exhibits a weak logarithmic change, which becomes negligible at power below $P_0$ (see Fig. \ref{fig:PowerOverview}).
The temperature dependence of the loss rate at the low power limit $P<P_0$ is shown in Fig. \ref{fig:LowPowerLimit}. It is assumed that the increase of losses at low temperatures is due to the presence of the TLSs with an energy gap corresponding to the energy of the resonator $h\nu_0$ in the active region of the CPWR. Therefore the loss rate can be splitted into two parts $\delta(T) = \delta_{\text{L}}+\delta_{\text{TLS}}(T)$, where $\delta_{\text{L}}$ describes the background loss rate and $\delta_{\text{TLS}}$ describes the losses induced by TLSs interaction. The background loss rate $\delta_{\text{L}}$ consists of coupling and radiation losses, and remaining losses in the superconductor. It is considered constant in the range of interest. The temperature dependence of $\delta_{\text{TLS}}(T)$ is given by the occupation difference of the energy levels of the TLSs. Therefore the loss rate can be characterized as follows:
\begin{equation}\label{eq:Model}
\delta(T) = \delta_{\text{L}} + F\alpha_{\text{TLS}} \cdot 4 \pi \nu_0 \tanh\left(\frac{h\nu_0}{2 k_{\text{B}} T}\right) \;.
\end{equation}
In this model the constant scattering parameter $\alpha_{\text{TLS}}=\pi D p_0^2 / 3\epsilon$ contains the density of states $D$ of the TLSs, their polarization $p$ and the dielectric constant $\epsilon$ of their host material (see Ref. \cite{Phillips1987}). The filling factor $F$ takes the distribution of the TLSs into account and is supposed to be constant for particular samples (see also Ref. \cite{Gao2008}).
From the fit of the experimental data (see Fig. \ref{fig:LowPowerLimit}), values $\delta_{\text{L}}$ and $F\alpha_{\text{TLS}}$ have been determined. It is important to note that the argument of the hyperbolic tangent is not fitted, but given by the energy gap $h\nu_0$ of the TLSs. The value of $F\alpha_{\text{TLS}}$ can be reconstructed from the temperature dependence of the fractional resonance frequency $\frac{\Delta \nu_0}{\nu}$ as well \cite{Gao2008}. The results obtained by both methods show excellent agreement (see Table \ref{tab:Table}). Surprisingly, the value of $F\alpha_{\text{TLS}}$ is of the same order for all investigated samples.
\begin{table}
\centering
\caption{\label{tab:Table}$F \alpha_{\text{TLS}}$ for the different samples obtained by fit to the temperature dependence of the loss rate and the resonance frequency. The values are obtained from the low power limit. The background loss rate $\delta_{\text{L}}$ and the exponent $\varphi$ of the power dependence is shown as well.}
\begin{ruledtabular}
\begin{tabularx}{\textwidth}{lllll}
Sample & $F \alpha_{\text{TLS}}$ (by $\delta$) & $F \alpha_{\text{TLS}}$ (by $\frac{\Delta \nu}{\nu_0}$)&$\delta_{\text{L}}$ (kHz)&$-\varphi$ \\ 
    Nb-Al$_2$O$_3$  & $2.6\cdot 10^{-6}$ & $2.4\cdot 10^{-6}$ & $31$ & $0.05$ \\

    Al-Al$_2$O$_3$  & $2.0\cdot 10^{-6}$ &  - & $25$& $0.03$\\
    
    Nb-Si  & $1.3\cdot 10^{-6}$ & $1.6\cdot 10^{-6}$ & $25$ & $0.16$ \\
\end{tabularx}
\end{ruledtabular}
\end{table}

\begin{figure}
\psfragscanon
 \psfrag{xl}[c]{Power at resonator input $P$ (W)}
 \psfrag{yl}[c]{$f(P)$}
 \psfrag{-18}{$10^{-18}$} \psfrag{-16}{$10^{-16}$} \psfrag{-14}{$10^{-14}$}\psfrag{-12}{$10^{-12}$}
 \psfrag{1}{1} \psfrag{0.9}{0.9} \psfrag{0.8}{0.8}  \psfrag{0.7}{0.7}\psfrag{0.6}{0.6}
 \psfrag{20 mK}{$20$ mK} \psfrag{200 mK}{$200$ mK} \psfrag{400 mK}{$400$ mK}
\includegraphics[width=0.90\linewidth]{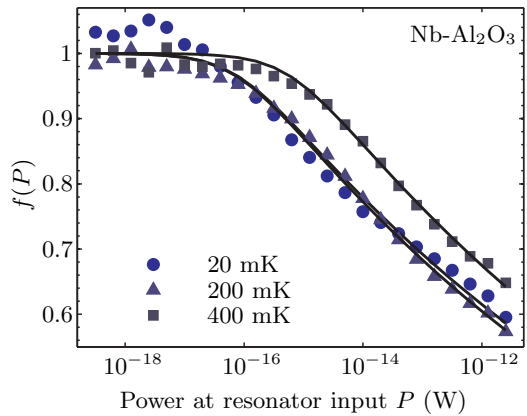}%
\caption{\label{fig:PowerNorm}The normalized power dependence of the loss rate at different temperatures for the Nb on sapphire sample. The low power limit is clearly visible. This normalization works for all samples and enables finding of $f(P)$. The exponent $\varphi$ of Eq. (\ref{eq:f}) is determined by fit (solid lines).}
\end{figure}

In order to describe the loss rate beyond the low power limit $P>P_0$ for different temperatures, we extend Eq. (\ref{eq:Model}) to
\begin{equation}\label{eq:Model1}
\delta(T,P) = \delta_{\text{L}} + F\alpha_{\text{TLS}} \cdot f(P) \cdot 4 \pi \nu_0 \tanh\left(\frac{h\nu_0}{2 k_{\text{B}} T}\right)\;,
\end{equation}
where the function $f(P)$ represents the loss rate's power dependence. In order to reconstruct $f(P)$ by making use of experimental data, we plot $\left(\delta(P)\vert_T-\delta_{\text{L}}\right)/F\alpha_{\text{TLS}}4\pi\nu_0\tanh\left(\frac{h\nu_0}{2 k_{\text{B}} T}\right)$ as a function of applied power (see Fig. \ref{fig:PowerNorm}).
The obtained function is constant ($f(P)$=1) for $P<P_0$ and has a linear dependence (in logarithmic scale) for $P_0<P$. Therefore this function can be described as:
\begin{equation}\label{eq:f}
f(P) = \left(1+\frac{P}{P_0} \right)^\varphi \;.
\end{equation}
Note that $P_0$ is slightly temperature dependent, and that the exponent $\varphi$ is temperature independent. Eq. {\ref{eq:f}} can be deduced for resonant interaction between the TLS and the photon field of the resonator, which gives a theoretical value $\varphi_{\text{th}}=-0.5$ \cite{Phillips1987}.
The values for $\varphi$ extracted from the measurement data differ from this theoretical value (see Table \ref{tab:Table}). 
This discrepancy reflects the fact that the power dependence is not defined by the resonant TLS interaction only,  but instead, depends also on substrate materials.

\begin{figure}
\psfragscanon
 \psfrag{y}[c]{Power $P$ (W)}
 \psfrag{z}[c]{Loss rate $\delta$}
 \psfrag{x}[c]{Temperature $T$}
 
 \psfrag{N}[Bl]{Nb-Al$_2$O$_3$}
  
 \psfrag{e}[Bl][bl][0.9]{$10^{-18}$} \psfrag{f}[Bl][bl][0.9]{$10^{-16}$} \psfrag{g}[Bl][bl][0.9]{$10^{-14}$}\psfrag{h}[Bl][bl][0.9]{$10^{-12}$}
 \psfrag{4}{40} \psfrag{6}{60} \psfrag{8}{80}  \psfrag{1}{100}   \psfrag{2}{120}
 \psfrag{a}{200} \psfrag{b}{400} \psfrag{c}{600}  \psfrag{d}{800}
 
\includegraphics[width=0.90\linewidth]{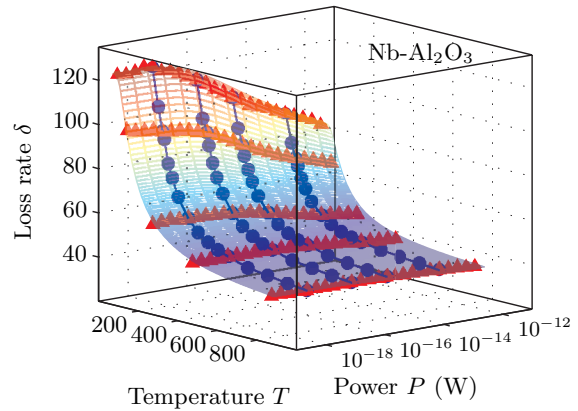}%
\caption{\label{fig:Complete}The complete description of the loss rate illustrated for the Nb on sapphire sample. The grid is generated using Eq. (\ref{eq:Model1}) with the parameters given in Table \ref{tab:Table}. Blue circles show measured data for the temperature dependence and the red triangles show measured data for the power dependence.}
\end{figure}

Using the extracted parameters, we are able to describe the temperature dependence at power levels $P>P_0$ and can therefore completely reconstruct the loss rate as a function of temperature and power (see Fig. \ref{fig:Complete}). However our approach is only valid for the TLS--dominant regime. The point where TLSs show a significant influence on the loss rate depends on their energy gap $h\nu_0$, which in our case corresponds to a temperature of $T=h\nu_0/k_{\text{B}}=120$ mK.  The influence of the TLSs on the loss rate becomes noticeable at $800$ mK, and significant at about $400$ mK, which corresponds to a $15 \; \%$ population difference. We note that the power range analyzed in this work reaches down to the low photon limit (at least $<10$ photons). 

In conclusion, we have shown that the intrinsic loss rate of a CPWR strongly depends on temperature which can be explained by the influence of TLSs in its environment.  The low power limit where the loss rate is solely temperature dependent was used to extract a measure for the number of the TLSs and their coupling to the photon field $F\alpha_{\text{TLS}}$. This parameter shows no significant difference between the material combinations used in this study. This material independence strengthens the evidence of a surface distribution of the TLSs \cite{Gao2008} and leads to the suggestion that the nature of the TLSs lies in general surface/interface states. 
Furthermore, the power dependence was phenomenologically described and a fully characterized model of the loss rate was developed. Hence the measurement data could be completely reconstructed.

\begin{acknowledgments}
We wish to thank S. Sauer, M. Grajcar, Ch. Kaiser, S. Tolpygo, M. Gershenson, V. Manucharyan and H. Paik for usefull discussions.
The authors gratefully acknowledge the financial support of the EU through the EuroSQIP project. EI acknowledges financial support from the Federal Agency on Science and Innovations of the Russian Federation under contract N 02.740.11.5067.
\end{acknowledgments}

\bibliography{}

\end{document}